\documentclass{aastex62}

\usepackage{xspace}
\newfont{\tensy}{cmsy10}
\newcommand{\chemical}[1]{{$\fontdimen16\tensy=3.0pt \fontdimen17\tensy=3.0pt \mathrm{#1}$}}
\newcommand{\Na} {\chemical{^{22}Na}\xspace}

\usepackage{enumitem}
\usepackage[colorinlistoftodos]{todonotes}

\usepackage{xspace}
\usepackage{isotope}
\newcommand{\Al} {\isotope[26]{Al}\xspace}
\newcommand{\Fe} {\isotope[60]{Fe}\xspace}

\usepackage{savesym}
\savesymbol{tablenum}
\usepackage{siunitx}
\restoresymbol{SIX}{tablenum}

\graphicspath{{./}{}}

\received{February 25, 2021}
\revised{some time in the future}
\accepted{some time in the future}
\submitjournal{ApJ}

\shorttitle{COSI: From Calibrations and Observations to All-sky Images}
\shortauthors{Zoglauer et al.}

\begin{document}

\title{COSI: From Calibrations and Observations to All-sky Images}

\correspondingauthor{Andreas Zoglauer}
\email{zoglauer@berkeley.edu}

\author{Andreas Zoglauer}
\affiliation{Space Sciences Laboratory, UC Berkeley, 7 Gauss Way, Berkeley, CA 94720, USA}

\author{Thomas Siegert}
\affiliation{Center for Astrophysics and Space Sciences, UC San Diego, 9500 Gilman Drive, La Jolla CA 92093, USA}

\author{Alexander Lowell}
\affiliation{Space Sciences Laboratory, UC Berkeley, 7 Gauss Way, Berkeley, CA 94720, USA}

\author{Brent Mochizuki}
\affiliation{Space Sciences Laboratory, UC Berkeley, 7 Gauss Way, Berkeley, CA 94720, USA}

\author{Carolyn Kierans}
\affiliation{NASA Goddard Space Flight Center, Greenbelt, MD 20771, USA}

\author{Clio Sleator}
\affiliation{U.S. Naval Research Laboratory, Washington, DC 20375, USA}

\author{Dieter H. Hartmann}
\affiliation{Department of Physics and Astronomy, Clemson University, Kinard Lab of Physics, Clemson, SC 29634, USA}

\author{Hadar Lazar}
\affiliation{Space Sciences Laboratory, UC Berkeley, 7 Gauss Way, Berkeley, CA 94720, USA}

\author{Hannah Gulick}
\affiliation{Space Sciences Laboratory, UC Berkeley, 7 Gauss Way, Berkeley, CA 94720, USA}

\author{Jacqueline Beechert}
\affiliation{Space Sciences Laboratory, UC Berkeley, 7 Gauss Way, Berkeley, CA 94720, USA}

\author{Jarred M. Roberts}
\affiliation{Center for Astrophysics and Space Sciences, UC San Diego, 9500 Gilman Drive, La Jolla CA 92093, USA}

\author{John A. Tomsick}
\affiliation{Space Sciences Laboratory, UC Berkeley, 7 Gauss Way, Berkeley, CA 94720, USA}

\author{Mark D. Leising}
\affiliation{Department of Physics and Astronomy, Clemson University, Kinard Lab of Physics, Clemson, SC 29634, USA}

\author{Nicholas Pellegrini}
\affiliation{Space Sciences Laboratory, UC Berkeley, 7 Gauss Way, Berkeley, CA 94720, USA}

\author{Steven E. Boggs}
\affiliation{Center for Astrophysics and Space Sciences, UC San Diego, 9500 Gilman Drive, La Jolla CA 92093, USA}
\affiliation{Space Sciences Laboratory, UC Berkeley, 7 Gauss Way, Berkeley, CA 94720, USA}

\author{Terri J. Brandt}
\affiliation{NASA Goddard Space Flight Center, Greenbelt, MD 20771, USA}



\begin{abstract}
The soft MeV gamma-ray sky, from a few hundred keV up to several MeV, is one of the least explored regions of the electromagnetic spectrum. 
The most promising technology to access this energy range is a telescope that uses Compton scattering to detect the gamma rays.
Going from the measured data to all-sky images ready for scientific interpretation, however, requires a well-understood detector setup and a multi-step data-analysis pipeline.
We have developed these capabilities for the Compton Spectrometer and Imager (COSI).
Starting with a deep understanding of the many intricacies of the Compton measurement process and the Compton data space, we developed the tools to perform simulations that match well with instrument calibrations and to reconstruct the gamma-ray path in the detector. 
Together with our work to create an adequate model of the measured background while in flight, we are able to perform spectral and polarization analysis, and create images of the gamma-ray sky.  
This will enable future telescopes to achieve a deeper understanding of the astrophysical processes that shape the gamma-ray sky from the sites of star formation (\Al map), to the history of core-collapse supernovae (e.g. \Fe map) and the distributions of positron annihilation (511-keV map) in our Galaxy.
\end{abstract}



\section{Introduction\label{sec:intro}}

The gamma-ray sky from a few hundred keV to several MeV is host to a multitude of phenomena, including the life cycle of matter in our Universe, the most energetic explosions (supernovae, mergers), and the most extreme environments such as pulsars and accreting black holes. Up to now, this energy range has not been very deeply explored, and a plethora of open questions remain, ranging from the origin of the 511-keV emission near the center of our Galaxy \citep[][]{Pranzos2011_511,Churazov2020:511}, to element creation during supernovae and mergers \citep{Isern2021:SN,Korobkin2020:Kilonova,Diehl2021:Nucleosynthesis}, and more.

The majority of these science goals require making images from the measured data showing point sources such as pulsars, binaries, and AGN, as well as diffuse emission from, for example, nucleosynthesis (e.g. \Al, \Fe) and positron annihilation (511-keV).
There exist several approaches for all-sky imaging in this energy range, such as coded-masks, rotating modulators, and Compton telescopes \citep[see for example][for overviews]{Schoenfelder2001,Diehl2018:Book}. 
Their effectiveness to identify and suppress background, the capability to measure polarization, their well-defined point-spread function, and their wide field-of-view give Compton telescopes a clear advantage over the other technologies in this energy regime to image point sources and diffuse emission. 
Therefore, Compton telescopes are at the center of this paper.

The path from the raw detector data to all-sky images using Compton telescopes has the following main steps: 

\begin{itemize}
\item Perform calibrations with sufficient statistics to accurately capture all features of the detector such as the shape of the energy response (photo peak as well as the Compton continuum), the shape and flux of the point spread function (peak, wings, and any features from incomplete absorption), the absolute flux normalization, and the scatter angle modulations due to polarization and detector geometry to a required accuracy.
\item Implement well-benchmarked Monte-Carlo simulations tuned to accurately reproduce these calibration measurements with all their detector features.
\item Reconstruct the path of the gamma-ray scatters within the detector from just position and energy measurements.
\item Create a multi-dimensional imaging response using Monte-Carlo simulations which sufficiently describes the point-spread function of the instrument.
\item Develop a model of the the on-orbit background either from observations or simulations as a function of location, time, and other environmental parameters.
\item Determine the source distribution on the sky using either an imaging approach such as Richardson-Lucy or maximum-entropy, or, alternatively, a model-fitting approach. All these methods need to take into account the all-sky scanning approach of modern Compton telescopes which means continuous slewing and maybe rocking of the instrument to achieve a close to flat exposure on the sky.
\end{itemize}

\begin{figure}
\centering
\includegraphics[width=0.6\textwidth]{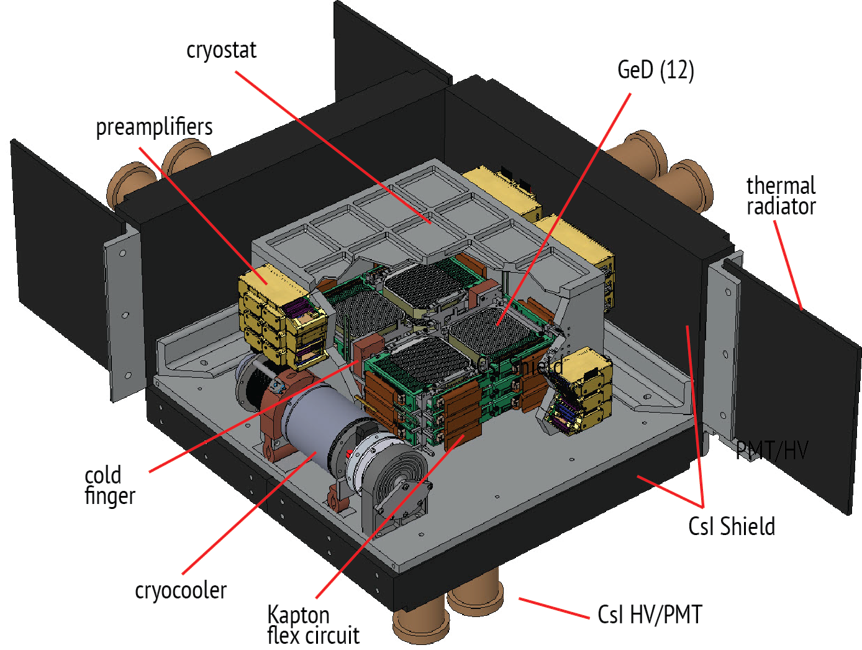}
\caption{Schematic of the COSI telescope showing the 12 germanium detectors (GeD) in the center enclosed in a cryostat which is cooled by the cryocooler. The cryostat is on five sides surrounded cesium iodide (CsI) shields, which are read out by high-voltage photomultipliers (HV/PMTs).}
\label{fig:COSISchematic}
\end{figure}

This paper is intended as an overview of how to perform the analysis while meeting these challenges and the lessons learned during the development of the data-analysis pipeline for COSI, the Compton Spectrometer and Imager (see Figure \ref{fig:COSISchematic} for a schematic).
COSI is a compact Compton telescope operating in the energy range from 0.2 to 5 MeV, capable of observing 25\% of the sky at any time.
COSI is a flexible telescope design that works on a variety of platforms. Here we concentrate on COSI-2016 \citep{Kierans2017:COSIBalloon}, with some references towards a planned satellite version, COSI-SMEX \citep{Tomsick2019:COSISMEX}.
COSI-2016, in the following referred to simply as COSI, consists of 12 high-purity, cross-strip germanium (Ge) detectors. The detectors have 37 orthogonal strips on each side with a 2~mm strip pitch and a 3~mm guard ring, resulting in a total detector size of 8$\times$8$\times$1.5~cm$^3$. The individual strips have an energy resolution of 2.5-4.0~keV FWHM and an average trigger threshold of 20 keV. The detectors are enclosed in a cryostat with a mechanical cryocooler keeping the germanium detectors at $\sim$ 83~K. The cryostat is surrounded on five sides by a 4-cm thick cesium iodide (CsI) shield which vetoes upward moving gamma rays and charged particles originating from cosmic-ray interactions with Earth's atmosphere. 

In 2016, COSI had a successful 46-day flight on a super-pressure balloon at an altitude of roughly 33.5~km launching from Wanaka, New Zealand and landing in Peru \citep{Kierans2017:COSIBalloon}. 
During this flight COSI observed the Galactic 511-keV emission \citep{Kierans2020:COSI511,Siegert2020:511}, gamma-ray bursts \citep{Lowell2017:GR160530A}, the Crab pulsar \citep{Sleator2019:Thesis}, black holes (Cyg X-1 and Cen A), relativistic electron precipitation events, and more. Our work with the COSI-2016 balloon data demonstrates the capability of this analysis pipeline for future flights on both balloon and satellite platforms.

\bigskip

Starting with a general introduction to Compton telescopes, the following sections will follow the data-analysis path with details on the steps outlined above, and ultimately demonstrate the pipeline by creating all-sky images from the COSI-2016 flight with several independent approaches.

\section{How Compton Telescopes Work}

\subsection{The principle of a Compton telescope}

Compton telescopes are named after Arthur Holly Compton, who discovered the scattering of X- and gamma rays off of electrons in 1923 and the relation between initial and final energy of the photon \citep{COM23}, later called the Compton equation (see Equation \ref{eqn:ComptonPhi}).
He received the Nobel prize in physics for this discovery in 1927.

\begin{figure}
\centering
\includegraphics[width=0.6\textwidth]{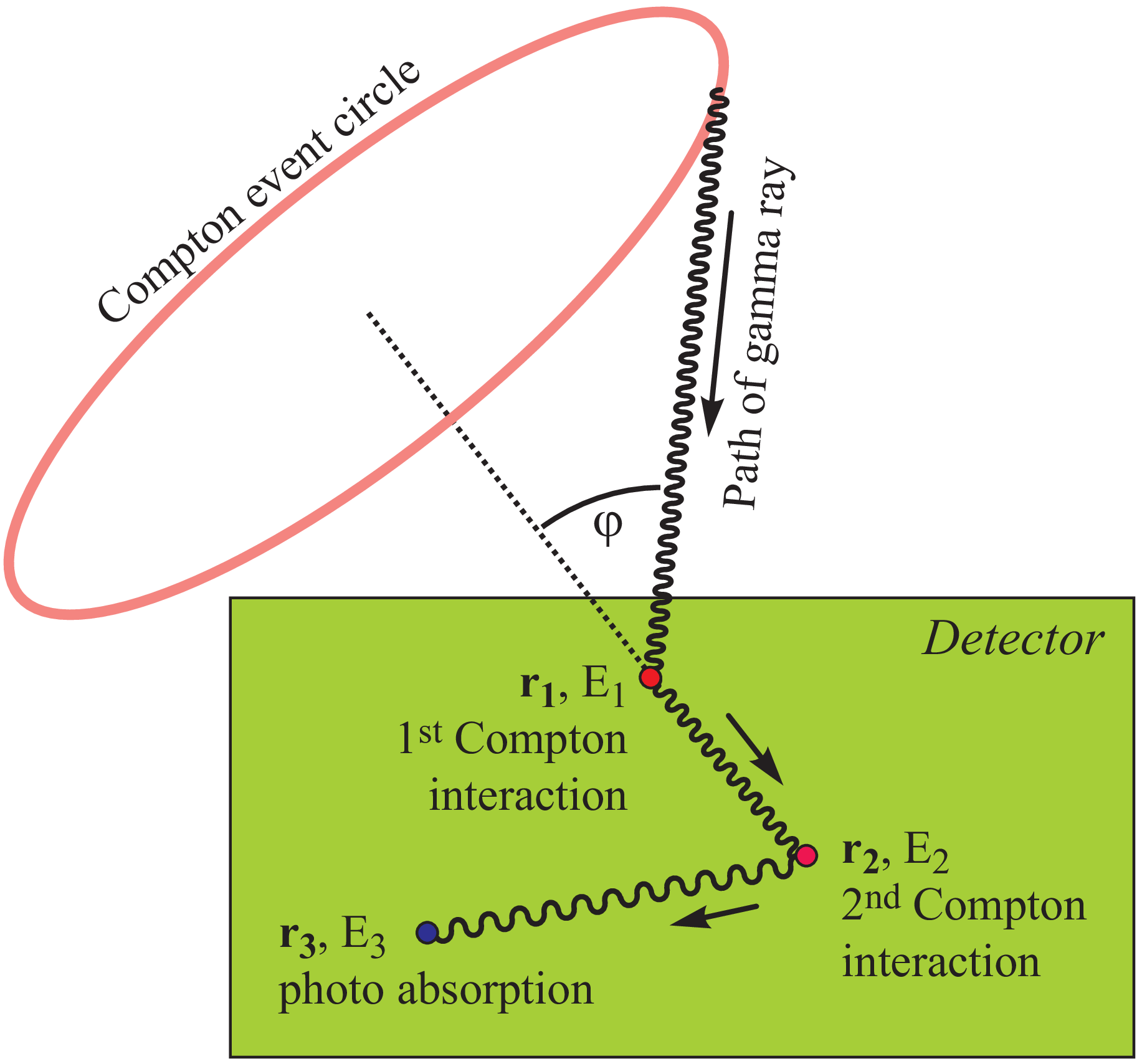}
\caption{The operating principle of a non-electron-tracking, compact Compton telescope such as COSI. The primary gamma ray undergoes one or more Compton interactions before it is ultimately stopped via a final photo absorption. The origin of the gamma ray can be restricted to a Compton event circle on the sky. The positions $\vec{r}_{1}$ and $\vec{r}_{2}$ determine the direction of the axis of the Compton event circle and the energies are used to determine the Compton scatter angle. }
\label{fig:ComptonPrinciple}
\end{figure}

The operating principle of a non-electron tracking Compton telescope can be found in Figure \ref{fig:ComptonPrinciple}: 
An incoming gamma ray with energy $E_i$ and momentum $\vec{p}_{i}$ scatters off of an electron in the active detector and transfers energy $E_e$ and momentum $\vec{p}_{e}$ to the electron. The scatter angle is called the Compton angle $\varphi$, and the corresponding electron is called the recoil electron. The gamma ray, which after the scatter has energy $E_g$ and momentum $\vec{p}_{g}$, continues on and undergoes either one or more Compton interactions or is absorbed in a final photo-effect interaction. The detector itself sees a set of energy deposits $E_n$ at locations $\vec{r}_n$ through ionization from the recoil or photo-electron. The true path of the gamma ray has to be reconstructed from just the energy and position information since the time-of-flight between the interactions is too small to be measured in compact Compton telescopes.

\subsection{Compton equation and cross section}

The scattering of a gamma ray off of an electron can be described in terms of conservation of energy and momentum of a photon and electron:

\begin{equation} 
E_{i} + E_{i,e} = E_{g} + E_{e} 
\label{ComptonEnergyConservation} 
\end{equation}

\begin{equation} 
\vec{p}_{i} + \vec{p}_{i, e} = \vec{p}_{g} + \vec{p}_{e} 
\label{ComptonMomentumConservation} 
\end{equation}

The initial energy $E_{i,e}$ and momentum $\vec{p}_{i, e}$ of a bound electron are not known beyond the boundaries set by atomic physics.
Therefore, it is commonly assumed that the electron is at rest, i.e.\ $E_{i,e} = 0$ and $\vec{p}_{i, e} = 0$.
The consequence of not knowing the initial kinematics of the bound electron is a fundamental physical limit for the angular resolution of a Compton telescope. This effect is called Doppler-broadening of the angular resolution (see also Section \ref{sec:Doppler} and \cite{Zoglauer2003:Doppler}).

Assuming we can measure the energy and direction of the scattered gamma ray ($E_{g}$, $\vec{e}_{g}$) and of the recoil electron ($E_{e}$, $\vec{e}_{e}$), we can use the relativistic energy-momentum relation $E_{e}^{rel} = \sqrt{E_{0}^{2} + p_{e}^{2}c^{2}} = E_{e}+E_{0}$ (with $E_{0}=m_{0}c^2$) and the relation between energy and momentum of photons $E_{g} = p_{g}c$ to derive the energy and direction of the initial photon:
\begin{equation} 
E_i = E_e+E_g
\end{equation}
\begin{equation}
\vec{e}_{i} = \frac{\sqrt{E_e^{2}+2E_eE_0}\vec{e}_{e}+E_{g}\vec{e}_{g}}{E_{e}+E_{g}} \\
\label{ComptonIdealOrigin}
\end{equation}

In the same way, the scatter angle $\varphi$ of the gamma ray can be determined. This is called the {\em Compton equation}:
\begin{equation}
\cos{\varphi} = 1 - \frac{E_{0}}{E_{g}} + \frac{E_{0}}{E_{g}+E_{e}}
\label{eqn:ComptonPhi}
\end{equation}

Compton telescopes consisting of large volume solid-state detectors, such as COSI, cannot determine the direction of the recoil electron $\vec{e}_{e}$. 
Due to this missing information, the origin probabilities of the gamma ray on the celestial sphere can only be restricted to a cone given by the above Compton-scatter angle and the direction of the scattered gamma ray (see Figure \ref{fig:ComptonPrinciple}, Compton event circle).

\bigskip

Five years after Compton's discovery, \cite{KleinNishina1928} derived the differential Compton cross section $\left( \frac{d\sigma}{d\Omega} \right)$ for unpolarized photons scattering off unbound electrons and then \cite{Nishina1928:PolComptonXsection} derived the differential cross-section for linearly polarized photons:

\begin{equation}
\left( \frac{d\sigma}{d\Omega} \right)_{Compton,\,unbound,\,polarized} = \frac{r_e^2}{2} \left( \frac{E_g}{E_i} \right)^2 \left( \frac{E_g}{E_i} + \frac{E_i}{E_g} - 2 \sin^2 \varphi \cos^2 \vartheta \right)
\label{equ:unboundpolarizedcrosssection}
\end{equation}
Here $\vartheta$ is the azimuthal or polar scatter angle.
Linearly polarized incoming gamma rays result in a cosine-shaped distribution in the azimuthal Compton scatter angle. This effect is most pronounced at lower energies and for Compton scatter angles around 90 degrees. See \cite{Lei1997} for a more in depth description of Compton polarimetry.

\subsection{The Point-Spread Function and the Compton Data Space \label{PSFCDS}}

Considering a normal camera (or even an X-ray focusing telescope), the direction of an incoming photon is translated into an x-y-position on the sensor. A point source will lead to a point-like peak on the sensor which is broadened due to imperfections in the optics. This is called the point-spread function (PSF). The space spanned by the x-y measurement positions is called the data space, and the sky is the image space. In this simple setup, the image is readily discernible in the x-y data space.

\begin{figure}
\centering
\includegraphics[width=\textwidth]{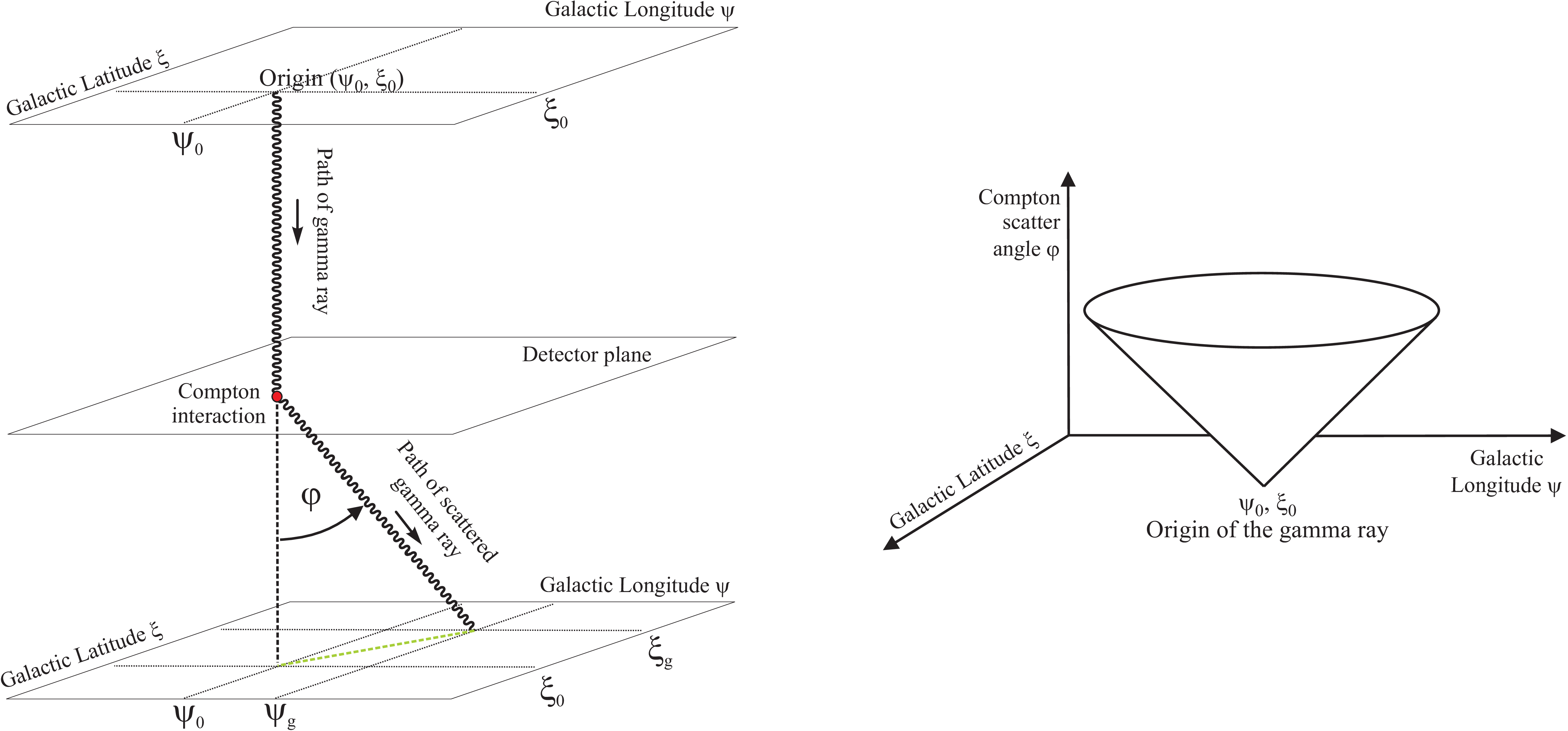}
\caption{Left: Schematic showing the connection between the Compton scattering process and the Compton data space: A gamma ray arrives at the detector from the direction $\psi_0$, $\xi_0$ in Galactic coordinates. It undergoes Compton scattering with the Compton scatter angle $\varphi$, and the scattered gamma ray continues on in direction $\psi_g$, $\xi_g$ in Galactic coordinates. Right: This information is then entered into the Compton data space spanned by the dimensions $\psi$, $\xi$, $\varphi$. In this data space, a point source creates a cone with a 90-degrees opening angle which points at the origin of the gamma rays ($\psi_0$, $\xi_0$).  }
\label{fig:PSF}
\end{figure}

A modern Compton telescope measures a set of positions with energies, i.e.\ the raw Compton data space consists of the dimensions $E_1, x_1, y_1, z_1, ..., E_N, x_N, y_N, z_N$, where $N$ is the number of interactions. 
The PSF is a complex hypersurface in this data space and an image is not easily discernible. 
However, this data space contains more information than is strictly necessary for data analysis and especially image reconstruction. 
For example, the key information that the interaction positions encode is contained in the first two interaction locations, representing the direction of the scattered gamma ray. 
The key information that the energies encode is the Compton scatter angle of the first Compton scatter.
The minimum data space for imaging therefore just contains these three dimensions: the direction of the scattered gamma ray in celestial coordinates such as Galactic longitude $\psi$ and latitude $\xi$ and the Compton scatter angle $\varphi$. 
To get images in celestial coordinates, all directions are converted from detector coordinates to celestial coordinates considering the pointing of the instrument. 
In this data space, a point source creates a cone with opening angle 90 degrees pointing at the origin of the gamma rays in the sky ($\psi_0$, $\xi_0$). 
This is a more intuitively discernible PSF of a Compton telescope. 
This data space is called the Compton data space (CDS) or sometimes, after the instrument it was first used for, the COMPTEL data space \citep[e.g.][]{Schonfelder1993}.
Figure \ref{fig:PSF} illustrates the connection between the Compton scattering process and the Compton data space.

According to the Klein-Nishina equation, the orientation of the scatter plane defined by the directions of the initial and the scattered gamma ray is random in the absence of polarization and geometry effects, resulting in the cone shape of the PSF.
The cone points at the source of the gamma rays since --- as the Compton scatter angle goes towards zero --- the direction of the scattered gamma ray becomes identical with the direction of the incoming gamma ray, which represents the source of the gamma rays (see Figure \ref{fig:PSF} left).
The cone has a 90-degree opening angle since the deviation of the scattered gamma-ray direction from the known origin direction is the Compton scatter angle, the same as the $\varphi$-axis in the CDS. This means the Compton scatter angle is encoded twice in the PSF, once derived from geometry and once derived from the kinematics. While this might seem redundant, it encodes important information about the response of the instrument, i.e.\ how accurately we can determine the Compton scatter angle with the detector due to the kinematics and position resolution.

The occupancy of this Compton cone is not equal everywhere. 
Along the Compton scatter angle axis, the summed occupancy is given by the probability that the gamma ray scatters by this Compton angle, which is determined by the Klein-Nishina cross-section.
In the $\psi$-$\xi$-plane, linear polarization of the incoming gamma ray would cause a cosine shaped amplitude modulation on the cone. 
Both distributions are modified again by the detector geometry, which might prefer or suppress certain scatter angles and directions.

This data space is a simplification, because it does not include all of the information contained in the raw data space which comprises all interaction locations and deposited energies. 
For example, the distance between the first two interactions influences the angular resolution. The uncertainties inherent to all measured locations influence the determination of the axis of the Compton cone, and this influence is largest for small distances between the interactions. 
As mentioned above, the occupancy of the cone along the Compton-scatter-angle axis is influenced by the Klein-Nishina equation, and thus smaller Compton scatter angles dominate at higher energies.
As a consequence, for the data analysis, in addition to the three dimensions in the default PSF, we ideally also want to include the measured energy and the distance between the first two interactions.

Concerning the relation of this data space with the image space, picking one ($\varphi$, $\psi$, $\xi$) element in data space, and determining which locations on the sky contribute to it, shows the classical Compton event circles in the sky (see Figure \ref{fig:ComptonPrinciple}), i.e.\ without knowing the direction of the recoil electron the gamma ray could have originated from any location on this Compton circle in the sky.

The width of the cone wall determines the angular resolution of a Compton telescope. 
The most frequently used proxy for the angular resolution is the ARM, the Angular Resolution Measure.
An example can be found in Figure \ref{Fig:ERComparison}.
The ARM is a one-dimensional projection of the Compton cone in the CDS;
it is a histogram of the smallest distance between a measured ($\psi$, $\xi$, $\varphi$)-value and the ideal Compton cone.
Since the ARM is a one-dimensional projection of the three-dimensional response, it has a few limitations.
Especially at smaller Compton scatter angles, which dominate at higher energies, it overpronounces the wings of the response, since there the counts in the wings spread out over a much larger area in the three-dimensional CDS than in the one-dimensional ARM.
In addition, any systematic deviations from the ideal Compton cone lead to a broadening of the ARM.
This, for example, happens for higher energy COSI events (above a few MeV). 
At these energies, the recoil electron can travel a few strips.
Since the start point cannot be determined with COSI, the center of energy is used as the location of the interaction.
This positional shift results in the direction of the scattered gamma ray to systematically being reconstructed a bit outside of the cone of the ideal PSF.
Since larger Compton scatter angles mean larger energy transfers to the recoil electron, this effect worsens with larger scatter angles.
Therefore, while in the CDS the width of the cone stays almost the same, the core of the ARM gets artificially broadened.
However, despite these limitations, the FWHM of the ARM is the most frequently used measure for the angular resolution of a Compton telescope.

Overall, the CDS is the space where all of the actual data analysis happens, not the image space with its overlapping Compton circles. 
In the 3-dimensional CDS, a point source is represented by a 2D cone and thus very uniquely confined. 
It thus holds a strong discriminating power, and the most sensitive analysis can only happen in this data space.

This is also the biggest advantage of Compton telescopes over their key competing telescope technology in this energy band, coded masks: while for a coded mask the PSF (the unique mask pattern for each direction) is distributed over the whole data space (in this case the detector plane), and therefore all measure background events contribute to the source-to-background ratio, for a Compton telescope only the events below the core of the PSF contribute. 
This improves the source-to-background ratio by typically a factor of 30--100 depending on the angular resolution.

\subsection{The broadening of the PSF\label{section:PSFBroadening}}

The PSF of a real-world Compton telescope is not a perfect cone. The cone wall is subject to broadening due to the uncertainties in the measurement process determining energies and positions, as well as due to physical limits. 
Below are all effects listed which influence the COSI-2016 PSF, roughly sorted from largest to smallest contribution to the shape of the core of the PSF.
However, as long as all of these effects are accurately accounted for in the simulations or added in the post-processing of the simulation data (see Section \ref{sec:DEE}, detector effect engine) and therefore are included in the simulated PSF's used for data analysis, no negative systematic effects are expected in the data analysis.

\subsubsection{Position measurement uncertainty}

COSI's cross-strip detectors have a fixed position resolution for each voxel (2$\times$2$\times$0.5 mm$^3$). This directly influences the accuracy with which the direction of the scattered gamma ray can be determined. This causes a certain jitter in the ($\psi$, $\xi$)-dimension of the PSF. 
Due to the smaller average distance between interactions at lower energies, and thus smaller lever arms, this effect is more pronounced at lower energies. 

For COSI, another effect is the travel range of the electrons in the germanium. The recoil electrons resulting from gamma rays with a several MeV can travel a few millimeters and thus can be measured across several strips. However, it is not possible to determine which strip has been initially hit. Therefore the average position is chosen as the interaction position, increasing the position uncertainty.

For COSI-2016, position uncertainties have the most influence on the width of the PSF and limit the angular resolution to a few degrees.

\subsubsection{Doppler broadening}
\label{sec:Doppler}

The standard Compton equation assumes that the gamma ray scatters off of an electron at rest. In reality the electron is bound to a germanium atom and has an unknown momentum at the moment of the scatter. This results in a slightly different direction and energy of the scattered gamma ray compared to what is expected from the standard Compton equation. For COSI, the additional broadening of the PSF is strongest at lowest energies and large scatter angles. This effect is called Doppler broadening and is the fundamental limit to the angular resolution that a Compton telescope can achieve \citep{Zoglauer2003:Doppler}. 
For germanium, this results in a roughly 1-degree (FWHM) resolution limit at 1\,MeV (depending on the exact event selections), and roughly 3\,degrees at 200\,keV.

\subsubsection{Energy measurement uncertainty}

The precision with which the deposited energies can be measured depend on detector material and the read-out electronics, and directly influence the calculated Compton scatter angle via Equation \ref{eqn:ComptonPhi}. 
Considering the 3D PSF, the energy resolution will cause a broadening of the PSF in the $\varphi$ direction. 
However, this is not a trivial relation. For example, the nearly constant energy resolution as a function of energy of COSI has a more substantial impact at small and large scatter angles \citep[see, e.g.,][chapter 2]{Zoglauer2005}.
Due to COSI-2016's excellent energy resolution, on average, this effect has only a minor influence on the broadening of the PSF.

\subsubsection{Lost interactions}

A realistic Compton detector contains passive material in addition to the active detectors and has a finite size. 
As a consequence, some of the Compton interactions will be in passive material and the finite size means that photons can escape the detector volume. 
If just the first interaction is in passive material, then the Compton sequence will look similar to a ``normal'' one which is completely contained in the detector, but the event will be off its true location in the Compton data space and will appear like background.
This is also the case if one or more of the middle or the final interactions are missing. 
However, at least some of these events can be rejected in the reconstruction process, since none of the possible Compton sequences will be compatible with the kinematics of Compton scattering.

\subsubsection{Charge sharing}

While the recoil electron is slowed down and stopped in the germanium, it creates electron-hole pairs which move to the opposing electrodes along the field lines. 
The distance travelled by the electron before it is stopped, along with charge diffusion and repulsion, cause a spread of the charge cloud. 
As a consequence, some interactions (more at higher energies) will be shared between multiple strips.
This has the effect of worsening the energy resolution since the energy measurement uncertainty of each strip impacts the total energy uncertainty. 
In addition, in some cases the energy shared on one of the strips might be below the read-out threshold leading to a energy loss up to the threshold energy. 
As consequence, the PSF will be broadened in the $\varphi$ direction.

\subsubsection{Reconstruction}

A critical step in the data analysis pipeline of Compton telescopes is determining the path of the gamma ray in the detector (see Section \ref{sec:ER}). 
In a few cases it will not be possible to find the correct path, due to measurement uncertainties, Doppler broadening, and since it is not always possible to determine the direction using Compton kinematics alone \citep[see][]{Zoglauer2005}.
If events with two interactions are incorrectly reconstructed, the flight path will simply be reversed. 
These events will create additional structures in the data space besides the Compton cone, and thus complicate the analysis.
For three or more interactions, the number of possible paths scales with the factorial of the number of interactions and any additional structures quickly blend into the noise far off the Compton cone.

\subsubsection{Excess interactions}

It is possible to have excess interactions. 
These can originate from random coincidences of two or more gamma rays.
Another possibility is a recoil electron leaving one germanium layer and hitting another one --- some of these events can be rejected since the location of the two interactions is right at the top and bottom edge of adjacent detectors. 
Finally, at least at higher energies, bremsstrahlung photons emitted from higher energy recoil electrons can result in stray hits.
If these event types are not identified, they will again appear as background in the CDS.

\subsubsection{Fluorescence}

Both photoelectric effect and Compton scattering kick an electron out of the germanium atom and leave a hole. The excited atom then relaxes via Auger effect (electron emission) or fluorescence (photon emission). 
The highest energy fluorescence photons (the K-alpha's and K-beta's around 10\,keV) can sometimes travel to the next strip, or, if the interaction happened close to the top if the detector, escape the detector and either interact in the next detector or in passive material. 
However, due to the low energy of these photons, any energy deposits are usually below the read-out threshold. This results in a small energy loss, and thus an additional broadening of the PSF in the $\varphi$-direction.

\bigskip

In summary, all these effects either influence the angular resolution or add background, and together ultimately reduce the sensitivity of the instrument. 
All of these effects are understood and accounted for the COSI data analysis pipeline.

\section{The Data-Analysis Pipeline \label{sec:pipeline}}

\begin{figure}
\centering
\includegraphics[width=0.86\textwidth]{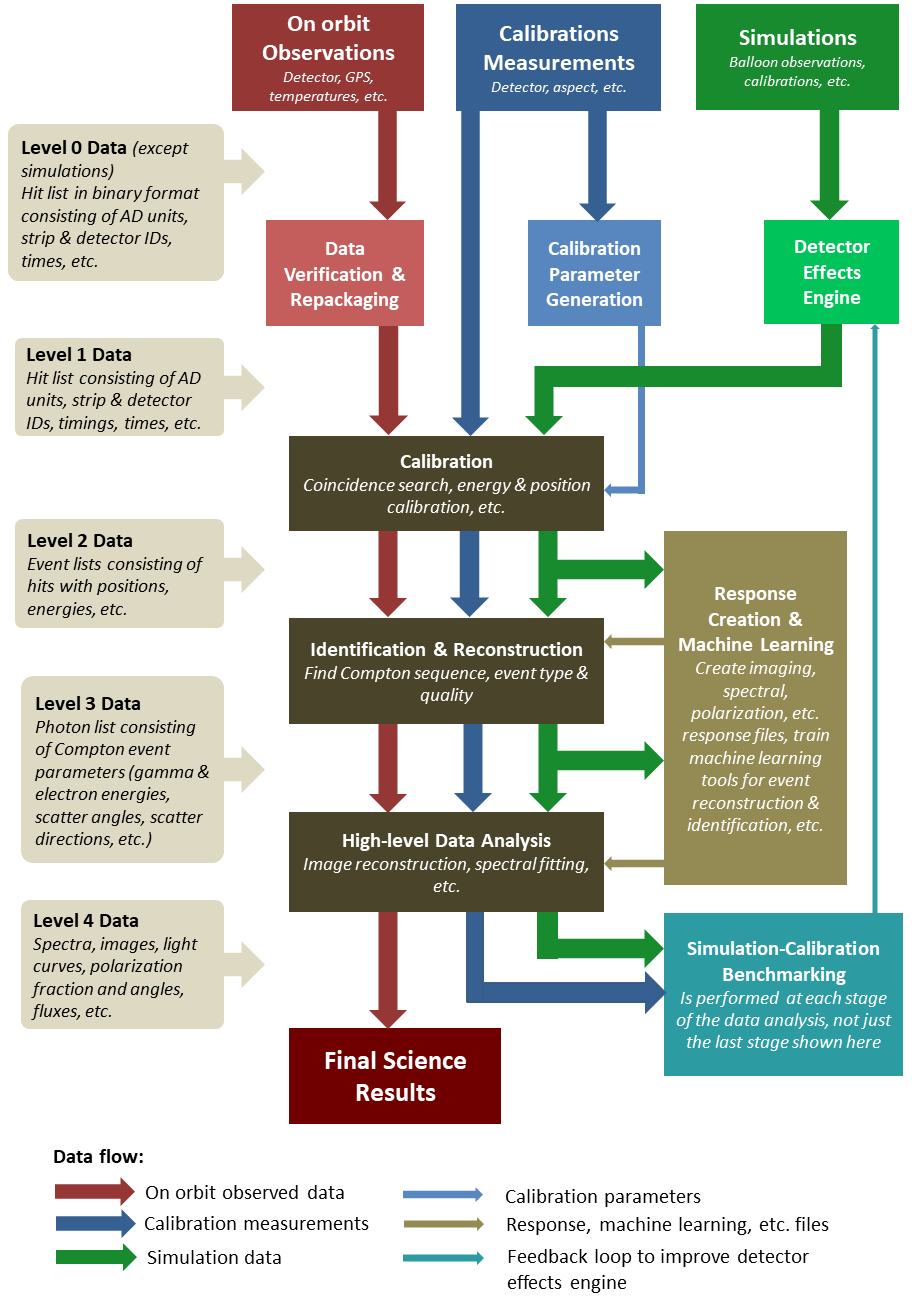}
\caption{The COSI data flow and data products from measurements, calibrations, and simulations to the final science results.}
\label{Fig:DataFlow}
\end{figure}

An overview of the COSI data analysis pipeline that is typical for a modern Compton telescope can be seen in Figure \ref{Fig:DataFlow}. 
The COSI pipeline consists of two main software components and a few smaller ones based on the former: Nuclearizer \citep[e.g.][]{Lowell2017:Thesis} for all calibration tasks, and MEGAlib \citep{Zoglauer2006:MEGAlib} for simulations and data analysis. 
The pipeline starts with three different input data streams stemming either from real on-orbit observation (red), ground calibrations (blue), or Monte-Carlo simulations (green). 
The simulations have to pass through a detector effects engine (DEE) first, which applies resolutions, thresholds etc., which ensures that simulations and measurements have identical characteristics. 
Subsequently, the three data streams pass through exactly the same pipeline, applying the calibration to measurements, reconstructing the events and identifying background, and finally high-level data analysis such as spectral and polarization analysis and all-sky imaging. 
Along this pipeline, the data are converted from the initial hits in detector units such as detector IDs, strip IDs, and analog-digital-converter units before the calibration, to event lists consisting of hits with positions and energies after the calibration, to photon lists containing, e.g., the parameters of the primary Compton interaction after the event reconstruction. 
The calibration measurements are used for two main tasks, to (1) determine the calibration parameters, e.g.,\ analog-digital-converter units of the detector's read-out electronics to energies, and to (2) benchmark the simulations, i.e.,\ ensuring that the simulations behave exactly like measurements in all aspects. 
The benchmarked simulations can then be used to create response files for imaging as well as other tasks such as to train neural networks for the event reconstruction. 
The following subsections describe the key elements of the pipeline in more details.

\subsection{Calibration measurements}

The first task of the calibration measurements is to create energy calibration files, dead strip lists, cross-talk correction files, threshold files, etc. Then, these will be used to calibrate and qualify the individual event data in the event calibration step.
The second task is to use the calibration measurements as a benchmark for simulations. This task sets the most stringent requirements on the calibration data taken, since it requires the collection of enough photons to show all effects causing the broadening of the response described in Section \ref{section:PSFBroadening} as a function of energy and field-of-view.

The following describes the key requirements to fulfill when developing a calibration plan for a Compton telescope.

\begin{itemize}
    \item The calibrations are best performed with (mono-energetic) lines from radioactive isotopes to disentangle full absorption (photo peak) from the Compton continuum. 
    In addition, the calibrations need to cover the whole energy range of the instrument. This is especially important at the higher energy edge, as the response starts to become distorted. In the case of COSI, this is caused by escaping gamma rays, as well as a longer range of the electrons in the germanium. This complicates the localization of the first interaction position and thus distorts the PSF. 
    Finally, the source activity needs to be well known. Since the calibrations directly inform the efficiency of the instrument, any uncertainty here would limit the accuracy of the absolute response normalization.
        
    \item Compton telescopes are inherently capable of observing gamma rays from all directions. Only geometry effects, shielding, and event selections can change that by blocking or rejecting some incoming directions, but even these effects need to be calibrated. As a consequence, the calibrations need to cover the complete field-of-view of the instrument. The key modifier which needs to be calibrated here are geometry effects, such as the change of the efficiency of the instrument as a function of incidence angle. For example for COSI, the effective area as a function of the incidence angle varies due to shield absorption and detector geometry. 
    Finally, COSI's anti-coincidence shields do not block all of the gamma rays. 
    Therefore, the transmission as function of incidence angle needs to be calibrated. 
    Our experience with COSI-2016 showed that at least $\sim$50 different incidence directions are required for a good calibration of the instrument.
    However, these different directions can be done with different isotopes.
    
    \item Simulations with at least a preliminary DEE are required to determine the minimum amount of photo-peak events needed to identify all features of the PSF. Usually it is not the core of the PSF that determines this amount but the wings of the response and any features in the wings in the Compton data space. See Figure \ref{Fig:ERComparison} for an illustration of the wings of the PSF in a one-dimensional projection.
    For COSI, the required number of calibration counts to fill the 4D data space of the response (PSF as a function of energy) is roughly 10 million events per line. However, not all calibration points need to have this number of triggers, since not all effects influencing the PSF depend on incidence angle. Typically, one would perform large calibrations on-axis for each source and then typically two more for off-axis angles, and the remainder with roughly 1/10 of the statistics.  
    
    \item None of these calibrations need to be performed with sources positioned far away from the detector. About one meter away (or just behind critical structures such as antennas) is enough. The main goal of these calibrations is to be a benchmark for the Monte-Carlo simulations. All the key far-field instrument parameters which describe the imaging performance such as effective area and angular resolution for a given incidence direction along with all response matrices can then be determined from the simulations.
    
    \item The source positioning must be accurate enough to not affect the angular resolution uncertainty or localization accuracy beyond the overall acceptable ground calibration uncertainty. For example, for COSI, 0.2~degree source position accuracy was acceptable given the instrument angular resolution of several degrees. 
    The extent of the source (i.e.\ the volume of the radioactive material) needs to be taken into account.
    
    \item Knowing the environment around the detector is key to reproducing the calibration with simulations. This includes the distance of the detector to the walls of the building, the material and density of the walls, any other objects close to the instrument of which gamma rays could scatter off (including the air with correct pressure and humidity), and any other radioactive sources close to the detector (e.g.\ Radon released into the air from the ground, K-40 included in concrete). In addition, good knowledge of the calibration source itself (material, amount, enclosure) is required, including the structure holding the source. All these need to be included in the simulation model for benchmarking.
    It is not enough to just know the environment, but part of the calibration plan should be regular background measurements, to understand the natural radioactivity around the source. 
    While this consumes some calibration time, we emphasize the need to define the wings of the PSF of a Compton telescopes accurately, especially because of the low number statistics there, and thus the relatively larger influence of scattered gamma rays and the natural background.

    \item Different objectives require different calibrations. For example, COSI's polarization capabilities need to be calibrated with either a polarized source or a normal radioactive source Compton-scattered off a secondary detector to create secondary polarization \citep{Lowell2017:Thesis}. 
    All the above considerations (e.g.\ energy range, field-of-view) are still valid for polarization calibrations. 
    In addition, for COSI's low energy calibration, sources need to be placed close to the corners of the detectors, so that the low energy gamma rays can reach each detector individually.
    
    \item Different instrument sub-systems require different calibrations. For example, COSI's shields and the guard rings surrounding COSI's cross-strip detectors will need their own calibration plan and are as critical for getting the data analysis correct as the main detector.
    
\end{itemize}

\subsection{Simulations and Benchmarking \label{sec:DEE}}

Monte-Carlo simulations are used to predict the performance of the instrument, to carry out trade-off studies for the instrument design, for observation planning, to compare observations with simulations, to train machine learning tools for event reconstruction, and to create response files for imaging, spectral, and polarization analysis. 
In order to accomplish these tasks, the simulations must be able to reproduce the measurements within a certain, acceptable margin. 
The process of matching simulations with calibration measurements is called benchmarking. The part of the software which makes the simulations resemble the actual measurements is the detector effects engine (DEE). 
COSI's simulations engine is cosima \citep{Zoglauer2009:cosima}. Cosima is part of MEGAlib and based upon Geant4 \citep{Agostinelli2003:Geant4,Allison2016:Geant4} and Geant4 is very well benchmarked in COSI's energy range.
These are the steps the simulations and DEE need to cover to achieve a good benchmarking:
\begin{itemize}
    \item A central element of the simulation and the DEE is the mass model of the telescope. This mass model should accurately reproduce the actual setup regarding volumes, their material (elemental and isotopic composition, density), and their placement. The closer the volumes are to the detector, the more accurate the mass model should be. Inaccuracies here will show up in the efficiencies of the photo-peaks, the Compton continuum, and also during simulations of instrumental activation which requires the correct isotopic composition. Concerning the benchmarking of calibration sources, the mass model also should contain the details of the calibration source and the structures which hold the calibration source, since scatters in this material will be directly visible in the measured spectra. In addition, the general environment such as air, walls, floors, and natural radioactivity need to be included. 
    
    \item In the case of simulations of actual on-orbit performance, the simulations have to include the full on-orbit background including all components such as cosmic photons and primary cosmic-ray electrons, positrons, protons, and ions. 
    If the instrument is close to Earth, it has to include the secondary radiation generated from cosmic-ray interactions with the atmosphere such as Albedo photons, protons, neutrons, electrons, and positrons. If the satellite moves through the radiation belts (e.g.\ the South Atlantic Anomaly), the interactions with those protons and electrons have to be included as well. 
    Finally, besides the primary particles, the delayed radioactive decays after a certain time in orbit (e.g.\ one year) resulting from proton, neutron, and ion activation have to be included. 
    Cosima is set up to perform all these tasks. 
    A detailed overview of the background model is presented in \cite{Cumani2019:Background}.
    
    \item The next step is to perform the actual Monte-Carlo simulation of the source with Cosima/Geant4. The simulations must include all the relevant physical processes the particles can undergo, ranging from photo effect to pair creation for gamma rays, from multiple scatters to bremsstrahlung for charged particles, from elastic to inelastic interactions for hadrons, and nuclear decay for nuclei. Cosima/Geant4 are fully set up for these simulations.
    
    \item The result of the simulation is an event list containing the locations of energy deposits in the detector without any simulated noise (energy resolutions, etc.) applied. These data need to be stored with enough accuracy so that in a second simulation step the charge transport in the detectors can be simulated.
    
    \item The next step is to apply the detector effects engine to the simulated data. For COSI, the DEE includes the simulation of charge transport of electrons and holes to the electrodes, the discretization of the locations into strips, handling energy resolution and read-out thresholds, position resolutions, analog-to-digital converter overflows, charge loss, coincidences, dead-time, and more. See \cite{Sleator2019:COSIDEE} for an overview of the COSI DEE.
    \footnote{Two limitations of the approach described in \cite{Sleator2019:COSIDEE} have been identified, regarding the simulation of the charge sharing and the COSI depth resolution making the performance shown in \cite{Sleator2019:COSIDEE} worse then it should be.} 
    
    \item Benchmarking the DEE is a complex, iterative, and thus time-consuming task. It involves comparing many aspects of the simulation data to the calibration data such as the individual spectra of individual strips, the combined spectra, the interaction locations in x, y, z, trigger rates, number of triggered strips per event and as a function of energy, the distances between the interactions, Compton scatter angles, the angular resolution, and ultimately the whole PSF. When differences show up, either individual effects in the DEE or the mass-model need to be adjusted. For example, differences in the photo-peak count rates in individual strips can be caused by differences in the coincidence and trigger system, issues with the mass model, wrong source intensity, incorrectly modeled charge sharing, etc. Differences in the width of the photo-peak usually just originate from the energy resolution. Tailing of the photo-peak, i.e.\ an asymmetry from the usually Gaussian shape extending to lower energies, originates from either charge loss between strips or charge sharing. Differences in the continuum can again be attributed to, for example, charge sharing, scatters in the environment or the source holder, or differences in the instrument mass model (e.g.\ missing mass). 
    Further details are presented in \cite{Sleator2019:COSIDEE}.
\end{itemize}

\subsection{Event calibration}

Event calibration is the first step which is identical for all three data streams in Figure \ref{Fig:DataFlow}. 
In this step, the data in detector units (strip and detector number, ADC units for energy, timing) are converted to interaction locations and measured energies in keV. For COSI, the event calibration is performed in the Nuclearizer tool \citep{Lowell2017:Thesis} and consists of these steps: 

\begin{itemize}
\item Coincidence: The time difference between hits is used to determine which hits belong to the same event.
\item Aspect determination and interpolation: The pointing information is determined more coarsely (once per second) compared to the number of triggered events. This step determines the pointing of the instrument in Galactic coordinates at the time the event is measured.
\item Energy calibration: The measured energy is converted from ADC units to keV for each triggered strip.
\item Charge-loss correction: Interactions spanning two or more neighboring strips lose some energy in the gaps between the strips. The energy loss follows a deterministic equation and thus can be corrected.
\item Cross-talk correction: Cross-talk is an artificial enhancement of the measured energy of one strip, when a neighboring strip also triggers. This effect can be calibrated and corrected in the COSI detectors.
\item Strip pairing: COSI's detectors are cross-strip detectors, i.e.\ a location can be determined from the x and y strip location. While this is trivial for one interaction, it is more complicated if there are two or more interactions in the detector. However, using the energy information, either a Greedy \citep[e.g.][]{Jungnickel:Greedy} or $\chi^2$-approach \citep{Zoglauer2005} can usually determine the interaction locations.
\item Depth calibration: When an interaction in the detector creates an electron-hole pair cloud, the electrons and holes drift to different electrodes. The relative arrival time between the electrons and holes allows for the calculation of the depth (the z-axis value) of the interaction in the detector. 
\item Position determination: The final step is to convert the interaction locations in the detector into the world coordinate system.
\end{itemize}

\subsection{Event Reconstruction \label{sec:ER}}

\begin{figure}
\centering
\includegraphics[width=0.6\textwidth]{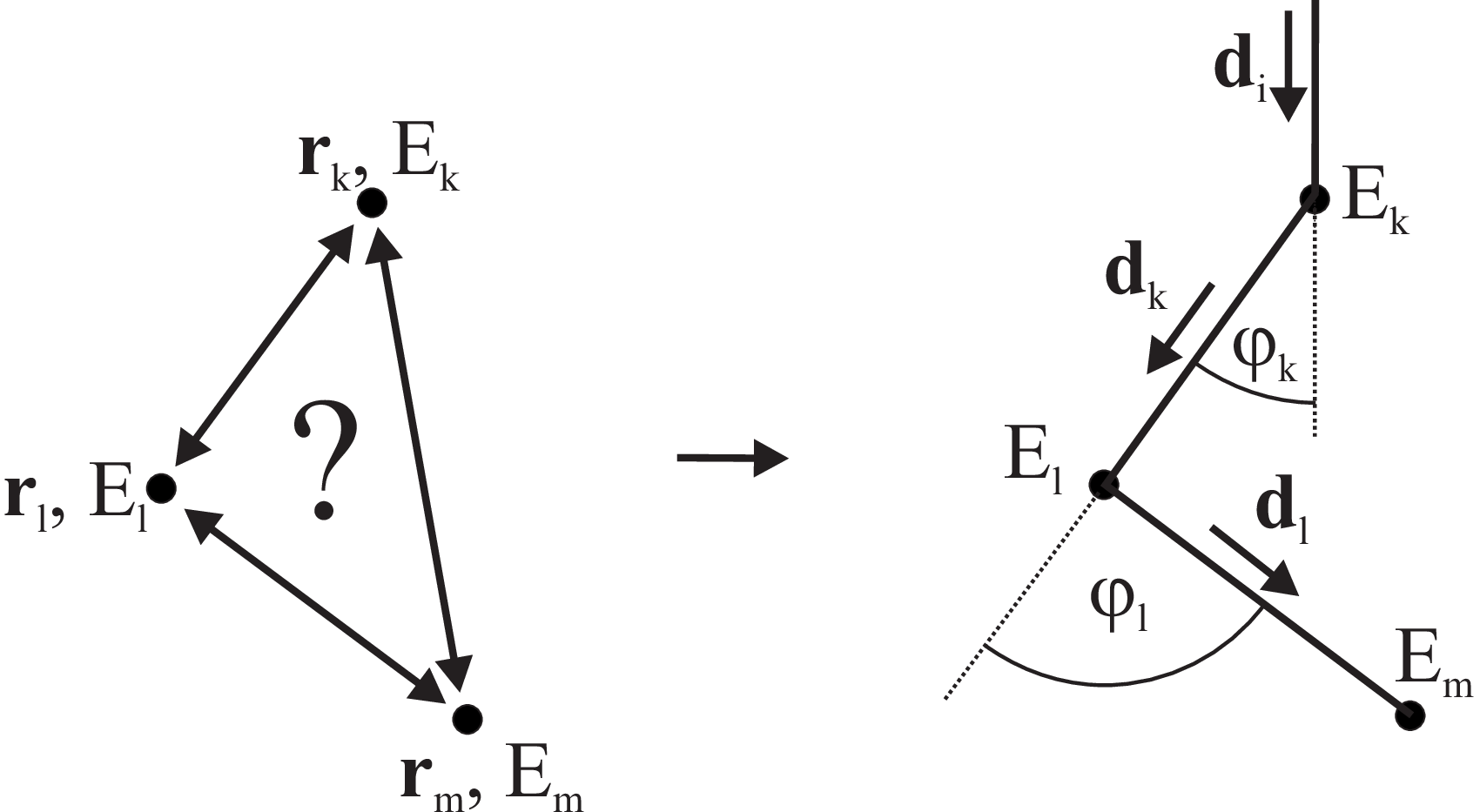}
\caption{The goal of event reconstruction: From a set of positions $\mathbf{r}_n$ and energies $E_n$ derive the known path of the gamma ray within the detector with known Compton scatter angles $\varphi_n$ and flight directions $\mathbf{d}_n$.}
\label{Fig:dPhi}
\end{figure}

The next step in the data pipeline is the event reconstruction, which converts the unsorted hits consisting of location and energy to the parameters of the initial Compton interaction (location, Compton scatter angle, etc.). This step includes the event pattern identification (is it a good Compton event?), Compton sequence determination (finding the path of the gamma ray in the detector), and background probability determination (did the gamma ray originate from the atmosphere below or from internal decay; did it escape?). Within MEGAlib, the ``Revan'' tool is responsible for the event reconstruction.

The critical step here is the Compton sequence reconstruction. The COSI detector is too compact and the time resolution in the germanium detectors is too coarse to determine the relative time of the individual interactions accurately enough to find the natural temporal sequence of the Compton interactions in the detector. Therefore COSI has to rely on the kinematics along with the interaction probabilities to determine the sequence of interactions.

The information which can be used includes:
\begin{itemize}
    \item The Klein-Nishina probability that a given Compton interaction happens with the detected Compton scatter angle given the measured energy of the event.
    \item The probability that the given distance between the interactions is observed given the material between the interactions and the energy of the travelling gamma ray. If this is the last segment of the Compton track, then the photo-absorption probability is used. Otherwise, it uses the Compton scatter probability.
    \item For middle interactions (see Figure \ref{Fig:dPhi}), the Compton scatter angle $\varphi_l$ can be calculated via kinematics (Compton equation using $E_l$ as energy of the recoil electron and $E_m$ as the energy of the scattered gamma ray) as well as geometry (difference between the incoming $\mathbf{d}_k$ and outgoing $\mathbf{d}_l$ gamma-ray direction at interaction $l$). The difference between those two angles should be zero within measurement uncertainties for the correct sequence.
\end{itemize}

Any event reconstruction approach now needs to look at all possible paths and determine the above information for each possible segment of the path. The number of possible paths is the factorial of the number of interactions, e.g.\ 3!=6 for 3 interactions. The task of the reconstruction approach is then to find the path which is most in agreement with the interaction physics, and therefore is most likely the correct path.

\begin{figure}
\centering
\includegraphics[width=0.6\textwidth]{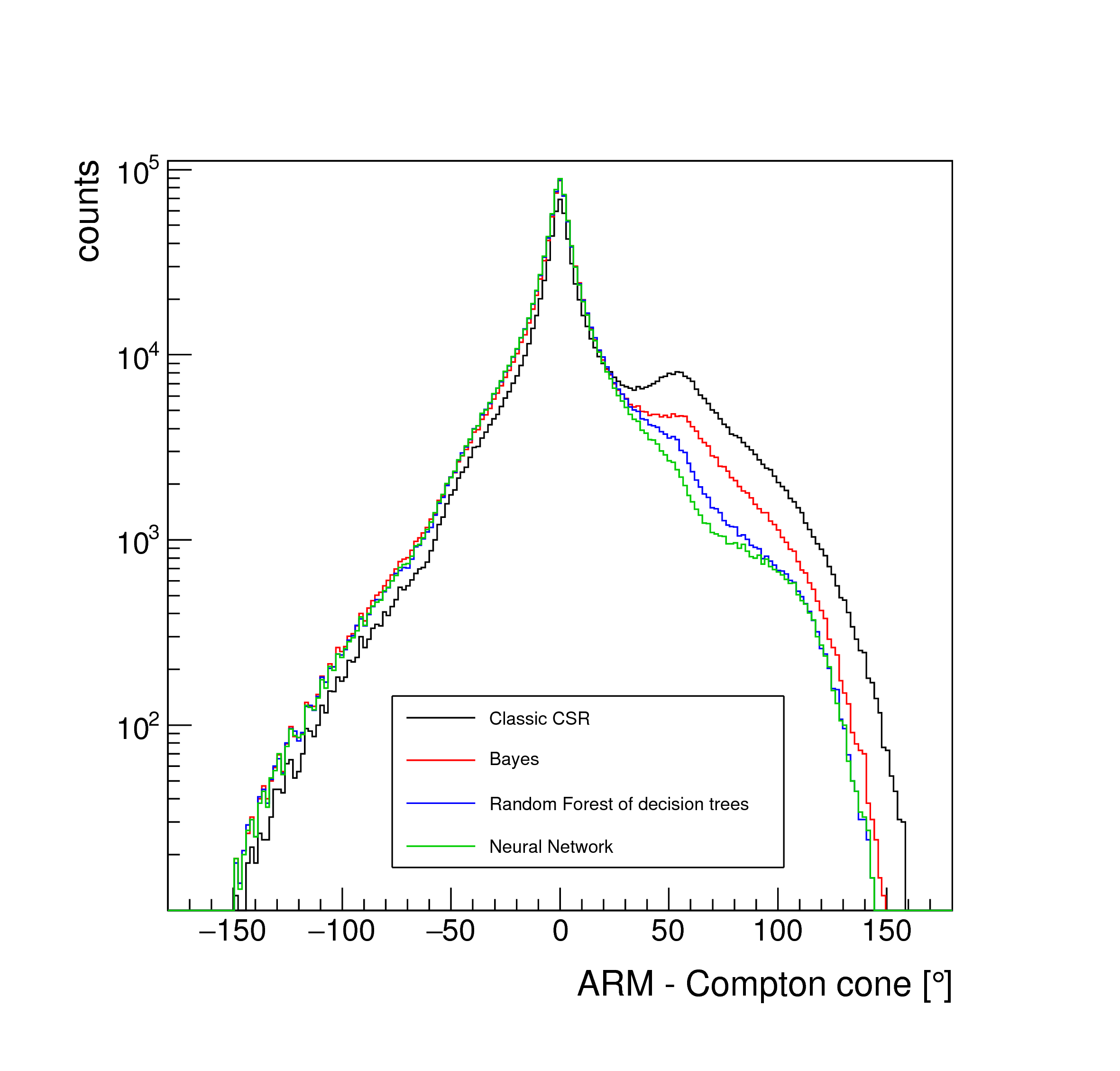}
\caption{Comparison of the different COSI event reconstruction approaches using the ARM as the performance metric. The count rate on the y-axis is plotted in logarithmic scale to accentuate the ``wings'' of the PSF, i.e., the counts accumulated far off the core of the PSF at 0 degrees. The correctly reconstructed events are in the peak around 0 degrees, wrongly reconstructed events accumulated at the bump between 30 and 120~degrees. The lower the bump, the more accurate is the event reconstruction approach. Therefore, the neural network approach is currently the best event reconstruction method.}
\label{Fig:ERComparison}
\end{figure}

To date, four unique approaches have been developed: the ``classic'' approach \citep[e.g.][]{Boggs2000:CSR, Zoglauer2005}, a Bayesian approach \citep[an older version is described in][]{Zoglauer2005, Zoglauer2007:BayesCSR}, and two machine learning approaches where one is based on a random forest of decision trees and the other is based on a shallow neural network \citep[an older version of this approach is described in][]{Zoglauer2007:NNCSR}. 
Figure \ref{Fig:ERComparison} shows a comparison of these approaches using an angular resolution measure (ARM) plot that includes data from an on-axis calibration measurement of COSI with a \Na source using the 511-keV line. 
The correctly reconstructed events show up around 0 degrees, and wrongly reconstructed events accumulate to the right in the bump. 
Currently, the most accurate event reconstruction approach is the neural-network method, since it produces the fewest number of incorrect event reconstructions, and therefore has the smallest bump at large angles.
Since the \Na source has an additional line at 1275 keV, the Compton continuum of this high-energy line contributes to the measured 511 keV photons used in this analysis. Therefore, the wings of the PSF are artificially broadened.

\subsection{Response and machine learning data sets creation}

Many of the approaches within the pipeline are driven by predefined data sets which describe how the instrument works. 
These include response files for spectral analysis and image reconstruction, as well as files containing the trained random forest and neural networks for event reconstruction.  
In general, these data sets are created from simulations where all necessary information about the initial parameters of the gamma ray and its interactions in the detector are stored.

For example, the training data sets for machine learning-based event reconstruction contain two data sets: one with the parameters of all the correct paths, and another one with all the incorrect paths. 
The information is derived from the simulations file which contains the sequence of interactions. 
Then the two data sets are used to train the neural network or the random forest to classify a given path as correct or incorrect.

Another example is the creation of response files for spectral analysis. This requires a large simulation of photons impinging isotropically on the detector with a flat spectrum covering the whole operating range of the detector. From this information a 4D matrix is created connecting the initial direction and energy of the gamma ray with the measured energy of the event.

An example of how to create the imaging response can be found in Section \ref{sec:ImagingResponse}.

\subsection{High-level data analysis}

The final step is the high level data analysis. This includes spectral analysis (\cite{Sleator2019:Thesis} presents results for the Crab and GRB160530A), polarization analysis (\cite{Lowell2017:Thesis} shows results for GRB160530A), and image reconstruction, to which the next chapter is dedicated.

\section{All-sky Imaging}

The measurement process of a Compton telescope can be described in the following way:
\begin{equation}
D(\psi, \xi, \varphi, E_m) = R(\psi, \xi, \varphi, E_m; \mu, \nu, E_i) \times I(\mu, \nu, E_i) + B(\psi, \xi, \varphi, E_m) 
\end{equation}
The sky distribution of gamma-ray emitting sources $I$, which is a function of the celestial sky coordinates (e.g. Galactic longitude $\mu$ and latitude $\nu$) and the energy of the gamma ray $E_i$, is convolved with the detector response $R$. Adding some background $B$ to the data space results in the measured data $D$. Here $\psi$ and $\xi$ are the direction of the scattered gamma ray in celestial coordinates (e.g. Galactic longitude and latitude), $\varphi$ is the Compton scatter angle, and $E_m$ is the measured energy.
Inferring the sky distribution $I$ from the measured data $D$ is an inverse problem. Since the detector measurement process includes a certain amount of randomness and due to the limited number of measured gamma rays, the problem is not directly invertible. As a consequence, the source distribution can only be retrieved using either model fitting approaches or iterative image reconstruction approaches.

For imaging, six key elements are required:
\begin{itemize}
    \item the infrastructure
    \item the observed or simulated data
    \item optimized event selections
    \item an accurate detector response
    \item a background model which ideally accounts for all individually varying background components
    \item one or more image reconstruction methods
\end{itemize}

The infrastructure is the MEGAlib toolkit and the observations and simulations are performed with COSI. The other elements are described below.

\subsection{Optimized event selections}

Some parts of the data space are highly contaminated with background and do not contain a significant amount of source photons. Eliminating all these events can improve the sensitivity of the resulting images.

For COSI, one of these selections is the so called Earth horizon cut. The Earth's atmosphere at COSI's flight altitude is extremely bright in gamma rays, and can dominate any source. To avoid this, we apply the Earth horizon cut, which eliminates all events whose Compton event circle dips below the 90-degrees zenith angle mark, which we consider being Earth's horizon.

Another key selection for COSI is a distance cut between the interactions in the detector. Events with a very small distance between first and second interaction have a very bad angular resolution (e.g.\ $\sim$55~degrees for $0.40\pm0.05$~cm distance at 662~keV). These events do not contribute very much to the imaging performance and are frequently incorrectly reconstructed. Therefore, COSI uses a standard 0.5~cm minimum distance cut.

\subsection{The Compton Imaging Response\label{sec:ImagingResponse}}

The instrument imaging response connects the image space to the Compton data space described in Section \ref{PSFCDS}. 
Assuming a binned image space spanning the celestial sphere in Galactic coordinates, the imaging response describes the probability that an event emitted in the given image bin (with the given energy and possibly polarization) is detected in a given data space bin. 
The simplest version suitable of nuclear-line imaging is a 5D matrix containing the Galactic latitude $\mu$ and longitude $\nu$ and connecting these to the Compton data space with the three dimensions Compton scatter angle $\varphi$ and the direction of the scattered gamma ray in Galactic latitude $\psi$ and longitude $\xi$. 
Picking one image bin and looking at the 3D Compton data space, one would see the familiar Compton cone described in Section \ref{PSFCDS}. On the other side, picking one data space bin and looking at the image space, one would see a Compton event circle centered on the direction of the scattered gamma ray with an opening angle corresponding to the Compton scatter angle. 
This corresponds to the Compton event circle in Figure \ref{fig:ComptonPrinciple}. 

Depending on the science objectives, different additional dimensions would be helpful for imaging.
For continuum imaging, an energy dimension is required (ideally one for the image space and one for the data space) to do full spatial-spectral deconvolution.
For polarization-resolved imaging, the polarization angle and amplitude could be added to the image space.
Furthermore, an additional dimension in the data space containing the distance between the first two interactions can help improve the performance for COSI, since the angular resolution is strongly influenced by this parameter --- the smaller the distance the worse the angular resolution.

Creating this response via simulations is the most computationally expensive task of the data analysis of a Compton telescope. For COSI, we use a 5-degree/$\sim$25-square-degree resolution of the 511-keV imaging response, resulting in 59,400 bins. These required roughly 3,300,000 NERSC hours on the Cori supercomputer.
We note that only $\sim 1 \pi$ steradian of these are occupied due to the field-of-view of the instrument.

As a remark, there is another approach to generate responses called ``list mode'' \citep[e.g.][]{Wilderman1998, Zoglauer2000, Zoglauer2011:Mimrec}, where the response is calculated from the measured parameters directly. While this method is well suited for terrestrial applications where the goal is to find single strong sources (e.g.\ environmental monitoring or even COSI calibrations), this method is not well suited for astrophysics, since it lacks the capability to model anything outside the core of the response and does not allow the determination of the absolute flux.  

The initial response as simulated is in detector coordinates. However, in order to perform imaging in Galactic coordinates, the response has to be converted into Galactic coordinates. COSI is a scanning instrument, i.e.\ it always points upward and does not stare at a fixed location. 
As a consequence, its pointing in Galactic coordinates constantly changes. Therefore, in order to get the response in Galactic coordinates, we have to follow the pointing in small time steps \cite[see][]{Siegert2020:511}, and rotate the response into the new coordinates and scale with time.
A special consideration for the COSI balloon flight is the atmospheric absorption of the gamma rays as a function of the zenith angle.
This reduction of efficiency has to be considered when transforming the response into Galactic coordinates.
This new response can then be directly used for imaging. It contains the probabilities that a gamma ray emitted from a certain position in the sky is detected in a certain data space bin within the total observation time. 

\subsection{The Background Model}

The low-energy gamma-ray regime is dominated by background --- a source to background ratio of 1 to 100 is not uncommon for sources at the detection limit, especially when the instrument is flown on a balloon platform. 
As a consequence, the retrieval of the source parameters (flux, spectral parameters, etc.) requires a robust background model. 
This can be developed in several different ways. For example, when looking at nuclear line emission from the Galactic disk, one can choose observations where the disk is not in the field-of-view. 
Alternatively, one can choose adjacent energy bands, e.g.\ 1820-1850~keV when looking at the 1809~keV line. Another option, assuming the background is well enough understood and is reproducible with Monte-Carlo simulations, is to simulate the individual background components. 
This would then allow for the influence of different components to be disentangled, for example, 511\,keV from internal beta-decays versus 511\,keV from atmospheric positron annihilation. 
Both components should behave differently in the data space as a function of altitude, latitude, and longitude for a balloon flight, or as a function of time after the last SAA passage (resulting in increased short term detector activation) for a satellite mission.

Similarly to the response, this initial background model will be in instrument coordinates and has to be rotated in small time steps along the pointing changes of the instrument into Galactic coordinates and scaled with time. 
The absolute normalization of the model will be part of the next step. However, if the model (or some of its components) vary consistently with some external parameter (e.g.\ balloon altitude, geomagnetic cutoff), then this scaling can be applied here.

\subsection{Creating images}

Now everything is in place for the final all-sky image creation.
An overview of all available approaches can be found in \cite{Frandes2016:Imaging}.
For space-based Compton telescopes, the most commonly used approaches are the Richardson-Lucy (RL) approach which is also known as Maximum-Likelihood Expectation-Maximization (ML-EM) \citep{Richardson1972, Lucy1974}, the Maximum-Entropy (ME) approach \citep{Strong1995:MEM}, the Multi-resolution Regularized Expectation Maximization (MREM) approach \citep{Knoedelseder1999:MREM}, and model fitting (MF) methods \citep[see e.g.][ for a direct application to COSI]{Siegert2020:511}.

The different methods have their advantages and disadvantages. The model fitting approaches, as the name says, are an easy way to compare different source distribution models, but can only find features which are part of the model or which can be composed of multiple copies of the model; MREM approaches can create a very smoothed solution with the risk of smoothing out weaker point sources or finer structures; RL approaches are better suited to find point sources but tend to quickly amplify noise; ME approaches are a middle ground between RL and MREM. 
All iterative methods (RL, MREM, ME) have the challenge that there is no well-defined way to determine when the iterations should stop. 
One approach is to look at the likelihood (RF, MREM) or the entropy (ME) at a given iteration, and stop the reconstruction when the changes fall below a preset level or start to oscillate.
Another way is to look at strong point sources (real ones or injected ones), and stop the reconstruction as soon as their resolution (e.g. FWHM) reaches the instrument resolution.

\subsection{Application to the COSI 2016 balloon flight}

\begin{figure}
\centering
  \begin{minipage}{\textwidth}
  \includegraphics[width=\textwidth,trim=300 100 300 100,clip] {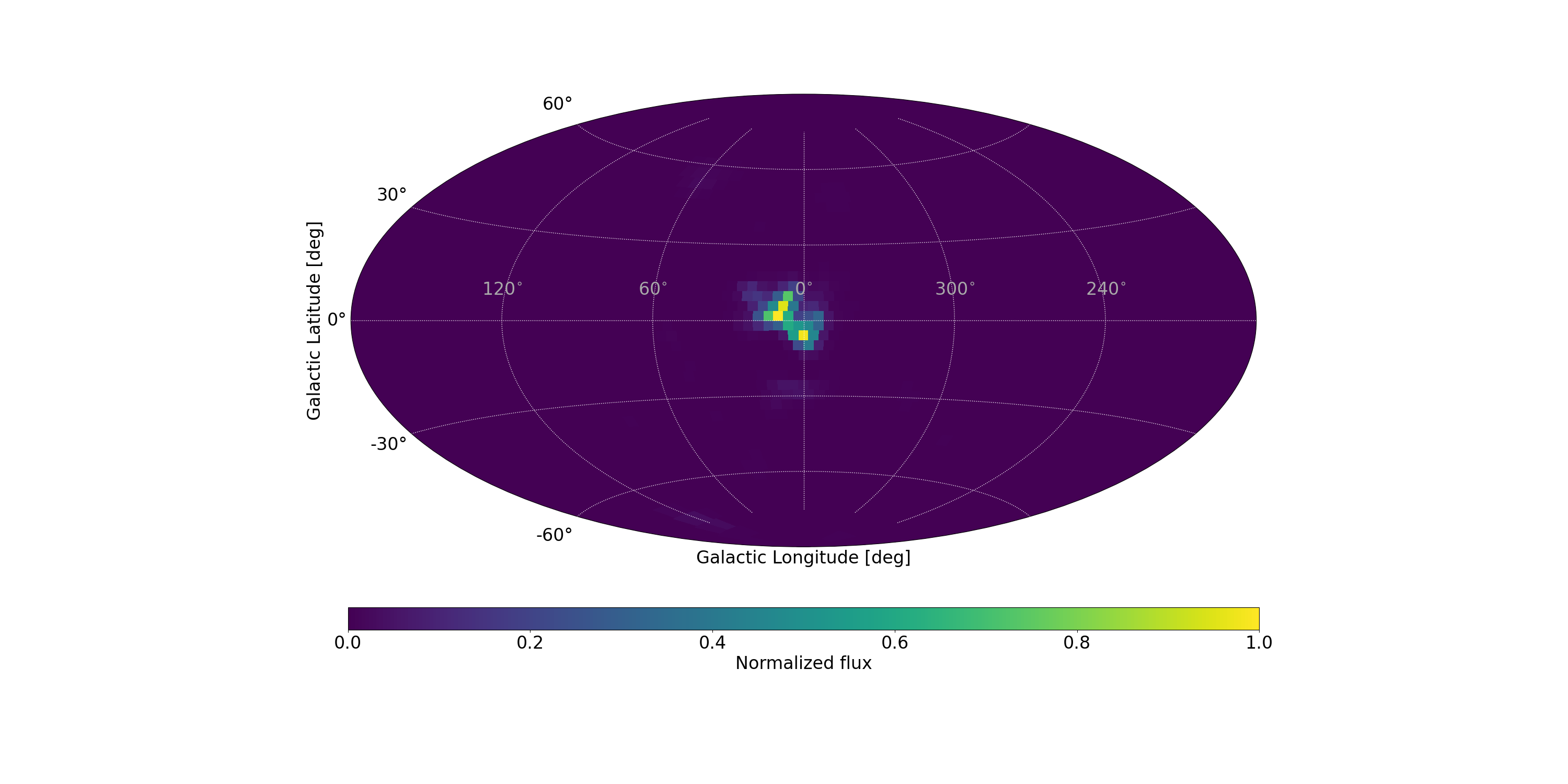}
  \end{minipage}
\caption{The 511-keV annihilation line image measured with COSI using 100\,iteration of a Maximum-Entropy deconvolution approach.}
\label{Fig:AnnihilationME}
\end{figure}

\begin{figure}
\centering
  \begin{minipage}{\textwidth}
  \includegraphics[trim=300 100 300 100 ,clip,width=\textwidth]{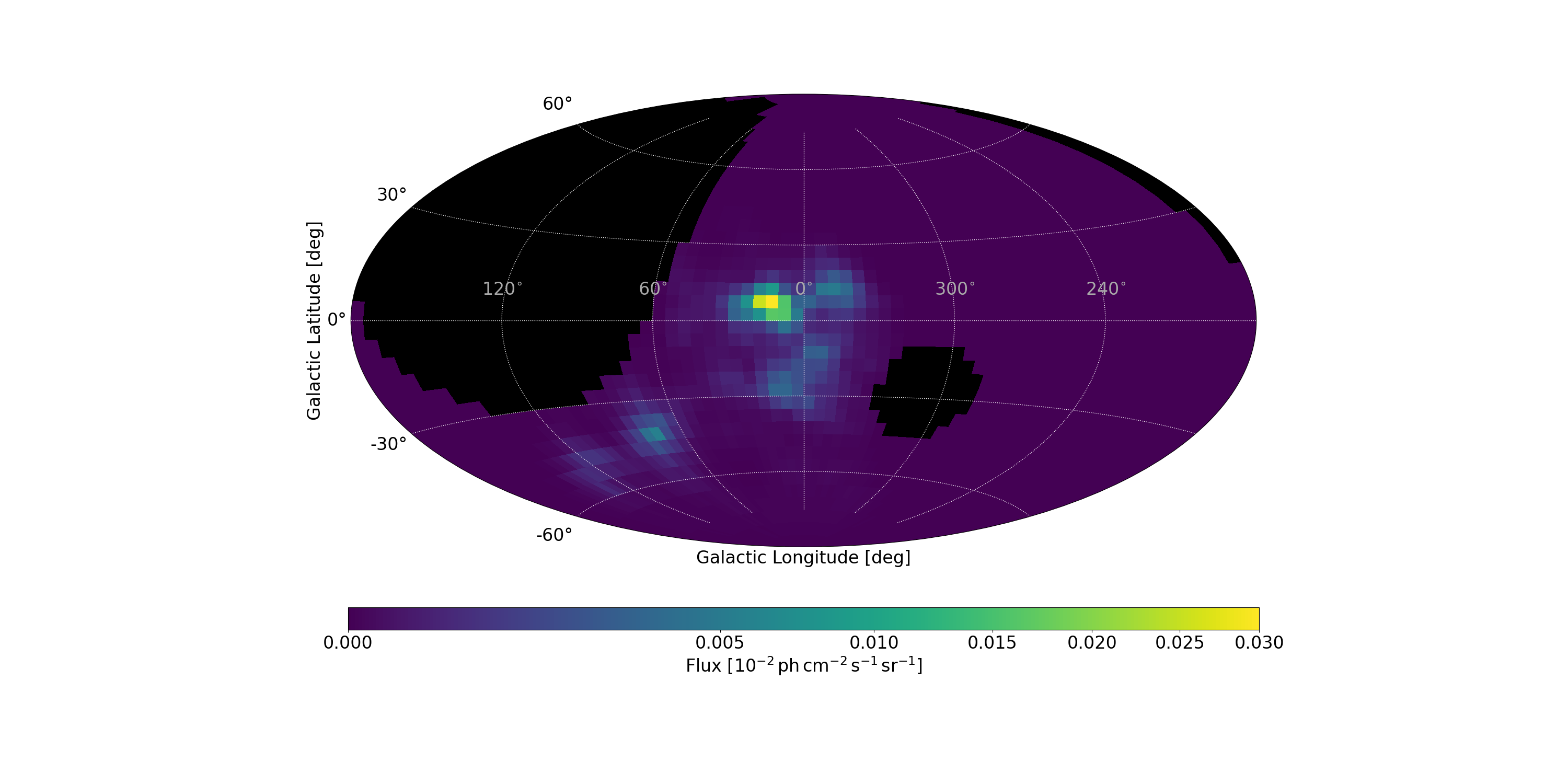}
  \end{minipage}
\caption{The 511-keV annihilation line image as measured with COSI using 26 iterations of an adapted Richardson-Lucy approach \citep[image adapted from][]{Siegert2020:511}. The black areas have no exposure and are excluded in the analysis.}
\label{Fig:AnnihilationRL}
\end{figure}

The most prominent all-sky source measured during the 2016 COSI balloon flight is the 511-keV annihilation line. 
As known from previous observations with SPI \citep[e.g.][]{Skinner2015:511}, the emission is concentrated near the Galactic center region, with much weaker emission from the Galactic disk.
With the limited balloon flight data, COSI achieved a 7-sigma spectral detection of this line itself \citep{Kierans2020:COSI511}.
However, spreading this 7-sigma signal in image space to generate an all-sky image results in large uncertainties.
Therefore, several approaches have been applied, including two image deconvolution approaches.
The methods differ not only in the algorithm used, but also the event selections, observation time selection, background modeling, and ultimately implementation (MEGAlib based on ROOT and C++ versus COSIpy, a Python3 tool, utilizing only parts of MEGAlib).
Figure \ref{Fig:AnnihilationME} shows the 511-keV\,emission after 100\,iterations of a Maximum-Entropy approach \citep{Strong1995:MEM, Hollis1992:MaximumEntropy}, and Figure \ref{Fig:AnnihilationRL} shows the 511-keV\,emission after 26\,iterations of a modified Richardson-Lucy approach. 
Details on the latter method and the event selections can be found in \cite{Siegert2020:511}.
The differences are mostly due to the limited statistics and different data selections. 
The images have in common that the emission is concentrated near the Galactic Center region and is extended, proving that COSI observed the Galactic 511 keV signal. 
For comparison, the angular resolution of COSI at 511\,keV is 6.6\,degrees \citep{Kierans2018:Thesis}.
The observed total flux is in agreement with SPI observations \citep[see][]{Siegert2020:511}.
However, none of the finer detailed features of these images are statistically significant.
This example shows that having multiple imaging approaches is especially helpful for sources close to the detection limit: features appearing in images created by different approaches are likely real.
A similar approach was taken for COMPTEL's analysis of the 1.8-MeV \Al map \citep{Knoedelseder1999:MREM} which used three different imaging methods.

\begin{figure}
\centering
  \begin{minipage}{\textwidth}
  \includegraphics[trim=300 100 300 100,clip,width=\textwidth] {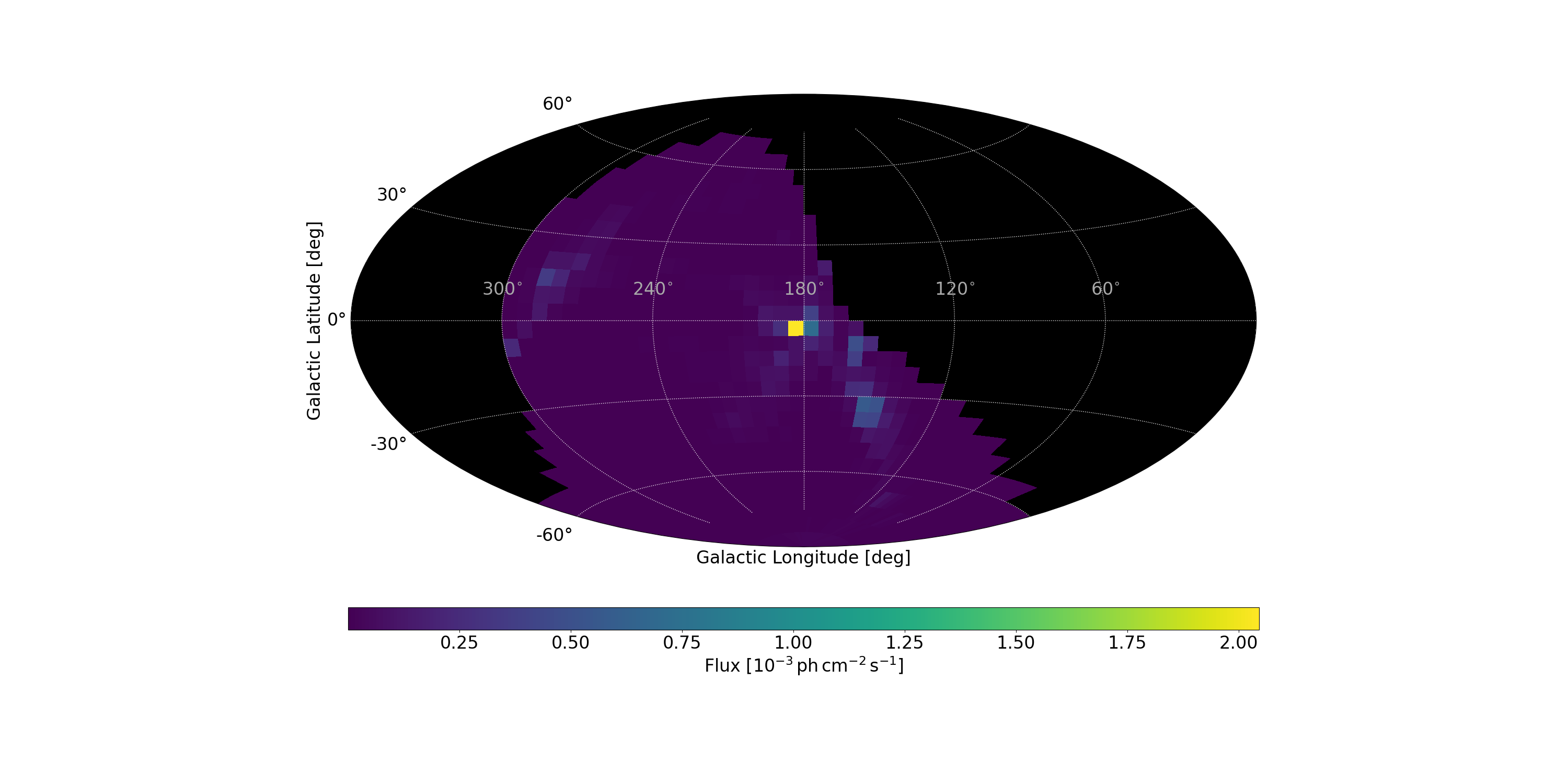}
  \end{minipage}
  \caption{Image of the Crab pulsar and nebula as seen with COSI in the 325-480\,keV energy band using 98\,iterations of an adapted Richardson-Lucy approach. The black areas have no exposure and are excluded in the analysis.}
\label{Fig:CrabImage}
\end{figure}

\begin{figure}
\centering
  \begin{minipage}{0.8\textwidth}
  \includegraphics[width=\textwidth]{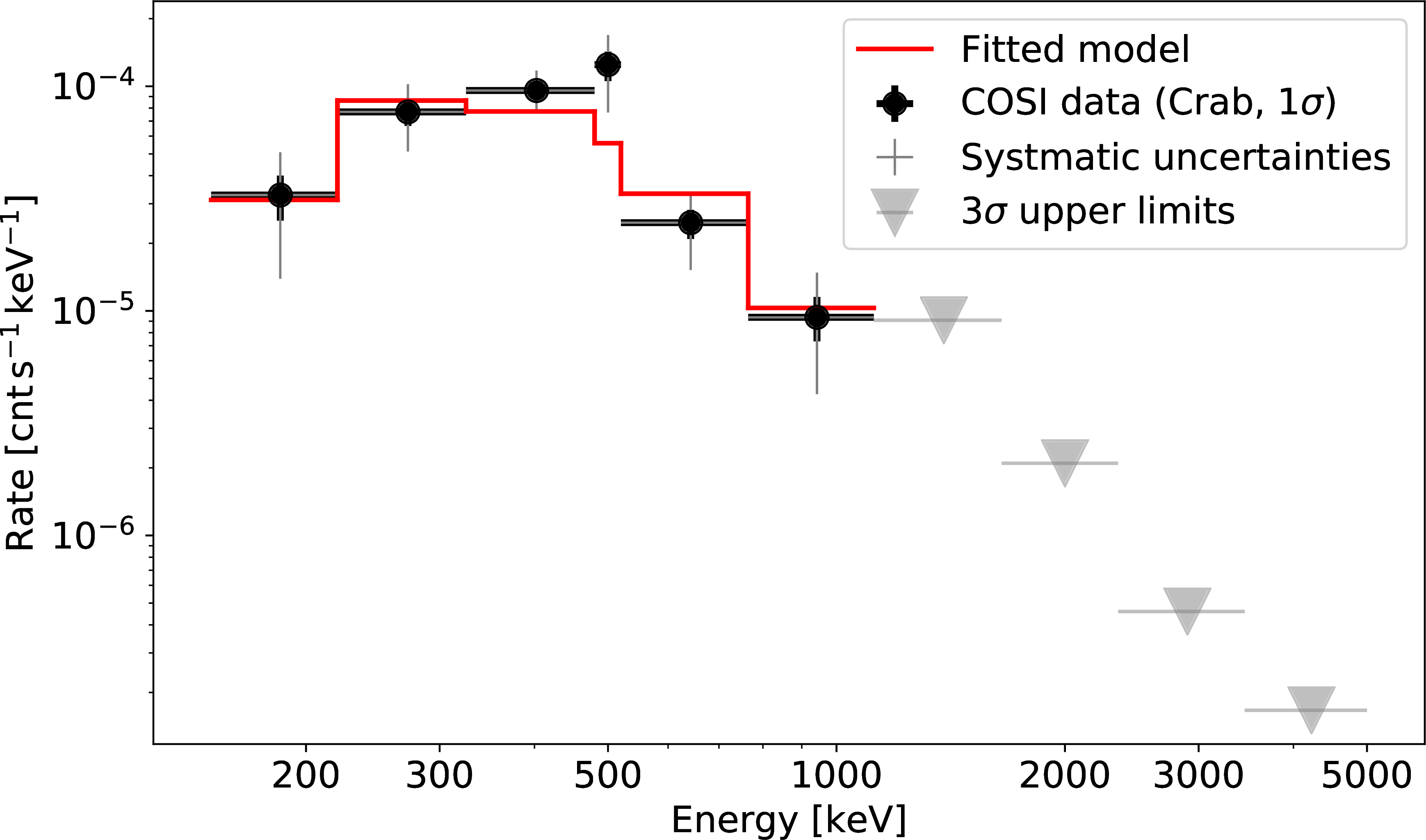}
  \end{minipage}
\caption{Measured Crab spectrum (black) which is in good agreement with the expected results: the measured power-law index is 2.5$\pm$0.4. }
\label{Fig:CrabSpectrum}
\end{figure}

Figure \ref{Fig:CrabImage} shows the same Richardson-Lucy approach applied to the observations of the Crab pulsar and nebula in the 325--480\,keV band --- COSI's efficiency peaks in this energy range.
After 98\,iterations, the Crab is clearly visible as the brightest spot.
The limited available simulation resources to create the seven dimensional continuum-imaging response --- the five dimensions of the CDS plus two energy dimensions --- limited the angular resolution in the image to six degrees.
For comparison, COSI's angular resolution in this energy band in $\sim$\,8\,degrees.
An independent analysis using the model-fitting approach from \cite{Siegert2020:511} shows that the Crab was detected at a significance of roughly 8\,sigma in this energy band.
Using all energy bands, the Crab is detected at 16\,sigma.
Since the Crab is located in the Galactic anti-center, it has the best visibility from the northern hemisphere. 
During the 2016 southern hemisphere flight, the Crab was only visible for a few days and at high zenith angles, when the flight path reached its northernmost point.
Therefore, the Crab gamma-ray emission observed by COSI passed through a significant amount of atmosphere. 
In the atmosphere, scattering of the gamma rays occur, which afterwards still can hit the detector but at a different angle and with lower energy. 
This results in the flux enhancements around the Crab in Figure \ref{Fig:CrabImage} beyond the spread due to the angular resolution.
It is possible to include this effect in the response, but this would require a significant amount of additional simulations of the complete atmosphere, which is beyond the scope of this paper.
The only atmospheric effect included in the response is the reduction of the photo-peak flux due to atmospheric absorption and scatters.
This effect is not an issue for nuclear-line images, since the selection on the 511-keV line automatically excludes scattered 511-keV photons. 
In addition, it is also not a problem for space missions, since there is no atmosphere.
All other features are not statistically significant and appear near the edges of the field-of-view with low statistics. 
The data used for the image include all times the Crab is less than 60\,degrees away from the zenith. 
As a side effect of this event selection, no other sources which have been detected with COSI are visible (Cen A, Cyg X-1), since they are too far away from the Crab.
The flux value ($3.9 \times 10^{-3}$\,ph/cm$^2$/s, central plus eight surrounding pixels since the angular resolution is larger than the pixel size) is within 5\% of the expected value \citep[$4.1 \times 10^{-3}$\,ph/cm$^2$/s, ][table 2, 2012--2019 values]{Jourdain:Crab} in this energy band.
Determining the uncertainty on this flux value would require detailed simulations of the scattering in the atmosphere and is beyond the scope of this paper.
Figure \ref{Fig:CrabSpectrum} shows the observed spectrum of the Crab with a fitted power law. 
The error bars are determined during the fit of the power-law spectrum of a point source at the Crab position and the background model to the data in the CDS.
The power-law index is 2.5$\pm$0.4 (68\% confidence uncertainty) which is in agreement with measurements from INTEGRAL \citep[2.32$\pm$0.02, ][]{Jourdain:Crab}.

\section{Conclusions}

This paper has shown the measurement, calibration, and reconstruction steps necessary to turn Compton telescope in-flight data into all-sky images. 
We have been able to identify and quantify the many inherent systematic effects influencing the COSI instrument and compact Compton telescopes in general. 
We have demonstrated a mature data-analysis pipeline ready for analysis of observations from the next gamma-ray astrophysics missions, including COSI-SMEX and other future missions.


\section{Acknowledgments}

We would like to thank the NASA Columbia Scientific Balloon Facility team for enabling the successful COSI-2016 balloon flight. The COSI-2016 balloon flight was funded by NASA APRA grant NNX14AC81G. Compton image reconstruction developments were supported through NASA grant NNX17AC84G. Machine-learning developments were funded by NASA grant 80NSSC19K0349. Thomas Siegert is supported by the German Research Society (DFG-Forschungsstipendium SI 2502/1-1). The COSI response simulations were performed on the Cori supercomputer (part of the National Energy Research Scientific Computing Center (NERSC), a U.S. Department of Energy Office of Science User Facility operated under Contract No. DE-AC02-05CH11231).

%

\vspace{5mm}
\facilities{COSI, CSBF, NERSC}


\software{MEGAlib \citep{Zoglauer2006:MEGAlib}, Geant4 \citep{Agostinelli2003:Geant4,Allison2016:Geant4}}

\bibliographystyle{aasjournal} 
\bibliography{references} 




\end{document}